
\input harvmac.tex
\def\thefig#1#2{\nfig\f#1{#2}}
\def\tiny{\scriptscriptstyle \rm}

\def\ofl{(\ell )}

\def\hef{{\sl He}$^4$}

\def\hes{{\sl He}$^6$}
\def\der#1{{d{#1}(\ell)\over d\ell}}
\def\dr{\delta\rho}

\def\pb{{\bf p}}
\def\rt{\right}
\def\lt{\left}
\def\half{{1\over 2}}
\def\grad{\hbox{${\bf\nabla}_{\!\scriptscriptstyle\perp}$}}
\def\qz{q_z}
\def\qp{q_{\scriptscriptstyle\perp}}
\def\bo#1{{\cal O}(#1)}
\def\dz{\hbox{$\partial_z$}}
\def\kbt{k_{\scriptscriptstyle\rm B}T}
\def\tb{\hbox{$\bf t$}}

\def\rb{{\hbox{$\bf r$}}}

\def\ddot{\!\cdot\!}

\hyphenation{Hibbs}
\Title{}{\vbox{\centerline{Directed Polymer Melts}\vskip2pt
\centerline{and}\vskip2pt\centerline{Quantum Critical Phenomena}}}

\centerline{Randall D. Kamien and David R. Nelson}
\bigskip\centerline{Lyman Laboratory of Physics}
\centerline{Harvard University}\centerline{Cambridge, MA 02138}

\vskip .3in
The statistical mechanics of directed line-like objects, such as
directed polymers in an external field,
strands of dipoles in both ferro- and electrorheological fluids, and
flux lines in high-$T_{\tiny C}$ superconductors bears a
close resemblance to the quantum mechanics of bosons in $2+1$ dimensions.
We show that single component and binary mixture critical phenomena in these
systems
are in the universality class of three dimensional
uniaxial dipolar ferromagnets and
ferroelectrics.  Our results also apply to films of
two superfluid species undergoing
phase separation well below their $\lambda$-points near $T=0$.
In the case of directed polymers and
electrorheological fluids
we analyze the effects of free ends occurring in the sample as well
as a novel directionally-dependent compressibility.


\Date{June 1992}

\newsec{Introduction and Summary}

There has recently been renewed interest in the analogy between
the classical statistical mechanics of
lines and the quantum mechanics of bosons in $2+1$ dimensions.
\ref\FEY{R.P.~Feynman, Phys. Rev. {\bf 91}: 1291 (1953); see also
R.P.~Feynman and A.R. Hibbs, {\sl Quantum Mechanics and Path Integrals}
(McGraw-Hill, New York, 1965) and R.P.~Feynman, {\sl Statistical Mechanics},
(Benjamin/Cummins, Reading, MA, 1972).}\nref\NS{D.R.~Nelson, Phys. Rev. Lett.
{\bf60}: 1973 (1988); D.R.~Nelson and H.S.~Seung, Phys. Rev. B
{\bf 39}: 9153 (1989); D.R.~Nelson and P.~Le Doussal, Phys. Rev. B {\bf 42}:
10113 (1990).}
\nref\LN{P.~Le Doussal and D.R.~Nelson, Europhys. Lett.
{\bf 15}: 161 (1991).}
\nref\KLN{R.D.~Kamien, P.~Le Doussal and
D.R.~Nelson, Phys. Rev. A {\bf 45}: 8727 (1992).}
\nref\N{For a review, see D.R.~Nelson, Physica A {\bf 177}: 220 (1991).}
Configurations of directed
lines can
be mapped onto the world lines of a collection of bosons in one
fewer dimension.  Examples include flux lines in high temperature
superconductors \NS , polymer nematics in a strong external field,
and ferro- and electrorheological
fluids \refs{\LN,\KLN}.

These systems can be modeled by
a free energy of the form \N
\eqn\eI{F=\int dz d^2\!r\,\left[{h\over 2}{\tb}^2 + {e\over 2}
\left(\grad\dr\right)^2 +{\alpha\over 2}\left(\dz\dr\right)^2
+{b\over 2}\dr^2 + {w\over 3!}\dr^3+{c\over 4!}\dr^4\right]}
The partition function
\eqn\eIa{Z=\int [d\tb] [d\dr]\,e^{-F/\kbt}}
is subject to the constraint
\eqn\eII{\dz\dr +\grad\!\ddot\tb=0,}
where $\dr(\rb,z)=\rho(\rb,z)-\rho_0$
is a cross-sectional areal density deviation from its average
value $\rho_0$ in a plane
perpendicular to $z$, and $\tb(\rb,z)$ is a tangent field in the same plane.
Given a microscopic description of directed polymers, via their trajectories
defined by single-valued functions
$\{\rb_j(z)\}$ along the $z$-axis, $\rho(\rb,z)$ and $\tb(\rb,z)$ are
defined by (see Figure 1)
\eqn\edensx{
\rho(\rb,z)=\sum_i\delta^2[\rb-\rb_i(z)]}
and
\eqn\emomx{\tb(\rb,z)=\sum_i{d\rb_i(z)\over dz}\delta^2[\rb-
\rb_i(z)].}

We assume that, as external parameters such as field strength, osmotic
pressure, temperature, solvent quality, {\sl etc.\ } are
varied, the polynomial part of \eI\ in $\dr$ exhibits a line of
first order phase transitions from a high to low density phase, terminating
in a second order phase transition.  Along the critical isochore, $w=0$, and
$b$
then changes sign at the mean field critical point.
Here and
throughout, we distinguish the $z$ direction as the direction
along which the lines are aligned, and denote by \grad\
the derivative perpendicular to the $z$-axis.  Additionally, we use
boldface to
denote vectors in the perpendicular plane.  In the case of
quantum mechanical bosons, we think of $\bf\hat z$ as a
time-like axis along which particle world lines are extended.  Planck's
constant
$\hbar$ plays the role of ``temperature'' in this system, $\beta\hbar$
is the sample thickness, and the coupling
$h$ is
the inverse ``mass density'' of the bosons.

The constraint \eII\ in the context of polymers
corresponds to the absence of polymer free ends \refs{\LN,\KLN} ,
that is, conservation
of polymer number as we would move up the $z$-axis.
In the context of quantum mechanical bosons, this constraint
merely sets \tb\ to be the canonical momentum density of the particles \KLN .
Models similar to \eI\ were first posed as Landau-Ginzburg type theories
\nref\DGII{P.G.~de Gennes, J. Phys. (Paris) Lett. {\bf 36L}: 55 (1975).}\nref
\MN{M.C.~Marchetti and D.R.~Nelson, Phys. Rev. B {\bf 42}: 9938 (1990);
Physica C {\bf 174}: 40 (1991).}\nref
\SB{J.V.~Selinger and R.F.~Bruinsma, Phys. Rev. A {\bf 43}: 2910 (1991).}\refs
{\NS,\DGII-\SB}
but were later shown to be a direct consequence of the boson mapping \KLN .

In fact, this model is equivalent to a three dimensional uniaxial
dipolar ferroelectric (or ferromagnet)
near its critical point, first studied in 1969 by
Larkin~and
\thefig{1}{Hydrodynamic volume averaging over directed lines in
$(d+1)$ dimensions which leads
to the coarse-grained density and tangent fields used in this paper.}
\noindent Khmel'nitski\u\i\ \ref\LK{
A.I.~Larkin and D.E.~Khmel'nitski\u\i , JETP {\bf29}: 1123 (1969);see also
A.~Aharony, Phys. Rev. B {\bf 8}: 3363 (1973); B {\bf 9}: 3946 (E) (1974).}.
To see this, we use the constraint \eII\ to solve for the longitudinal part
of $\tb$.  Upon integrating out the transverse part of $\tb$ we obtain
an effective free energy
\eqn\eIII{F_{\hbox{\rm eff}}=\half\int {d\qz\over 2\pi}{d^2\!\qp\over
(2\pi )^2}
 \,\left( h{\qz^2\over\qp^2}
+ e\qp^2 + \alpha\qz^2+b\right)\vert\dr(\qp,\qz)\vert^2+{c\over 4!}
\int dz d^2\!r\, (\dr)^4\; ,}
where the quadratic part has been written in Fourier space.
Here and henceforth, we restrict our attention to the critical isochore, and
set $w=0$.  Equation \eIII\ is
the familiar form of the Landau theory for a uniaxial Ising
ferromagnet or ferroelectric with a mean field critical point at
$b=0$ \LK .

A model very similar to \eI\ applies to polymer nematics in which the
external field is turned off, so that the alignment represents
{\sl spontaneously} broken symmetry instead of being externally imposed
\refs{\LN,\KLN,\DGII,\SB} .  In this case, $\tb(\rb,z)$ represents
a nematic director field and the term $\half h\tb^2$ is replaced by
the usual nematic gradient free energy
\ref\DGIII{P.G.~de Gennes, Physics of Liquid Crystals (Oxford University,
London,
1974).}.  Simple power counting arguments for this model suggest that
nonclassical critical behavior results below $(3+1)$-dimensions.  A complete
treatment, however, would require a discussion of the effects of hairpin
turns in the polymer configurations \KLN .  We restrict our attention here to
polymer nematics in a strong external field, and to length
scales less than the spacing
between hairpins, to avoid these complications.

\subsec{Electrorheological Fluids and the Effect of Free Ends}
The effective attraction for polymers in a nematic solvent is
similar to the fluctuation induced attraction of strings
of colloidal spheres discussed by Halsey and Toor \ref\HTI{T.C.~Halsey
and W.~Toor, J. Stat. Phys. {\bf 61}: 1257 (1990).}.  This attractive
part of the interchain potential in electrorheological fluids can
be offset by a hard-core repulsion at short distances,
and may be adjusted by varying the dielectric constant of the
solvent.  Since an electric field is required
to create directed strings of dielectric spheres,
hairpin configurations will be greatly suppressed in electrorheological
fluids, as there is no longer an up-down symmetry.  We then expect
that \eI\ is especially apt for these systems, once free
ends are taken into account.  Halsey and Toor \ref\HTII
{T.C.~Halsey and W.~Toor, Phys. Rev. Lett. {\bf 65}: 2820 (1990).},
by considering the bulk energetics of the system,
have argued that at sufficiently large fields, the electrorheological
strings will phase separate, leading to two-phase
coexistence.  The critical point discussed here would
then describe a transition to a uniform density.

To study this point further, consider the free
energy at fixed potential due to the electric field
\eqn\eVII{U=-{1\over 8\pi}\int dV\,\epsilon[n(\vec x)]{\vec E}^2(\vec x)}
where $n$ is the full three-dimensional colloid density,
and $\epsilon(n)$ is the
effective dielectric constant.  There is an overall minus sign
because this is the energy at fixed $\vec E$.
Since $\epsilon''(n)>0$ \ref\END{
For a system of spheres arranged in a cubical lattice or amorphously,
the Clausius-Mosotti relation holds, so $\epsilon(n)=(1+2v\gamma n)/
(1-v\gamma n)$, where $\gamma$ is the molecular polarizability and
$v$ is the molecular volume.  In this case $\epsilon''(n)$ is explicitly
positive.  A virial expansion for the bulk dielectric constant
in terms of the density can be done in general.  The first virial term
leads to the Clausius-Mosotti relation, and the second term is generally
quite small.  Thus $\epsilon''(n)$ is expected to be positive
even if the spheres are not arranged either cubically or randomly.
See G.C.~Maitland, M.~Rigby, E.B.~Smith and W.~Wakeham, {\sl Intermolecular
Forces}, (Clarendon Press, Oxford, 1981).}, it is energetically
favorable to phase separate into dense and dilute regions.  However, at
finite temperature, configurational entropy plays a role.
Certainly, at fixed electric field, the temperature can be
increased until entropy dominates and there is only one
uniform phase of directed lines.  However, entropy also favors shorter
chains at such temperatures, which could obscure the entanglement effects
which are the main focus of this paper (see below).

Recent numerical simulations by Tao \ref\T{R.~Tao, Southern Illinois
University Preprint, (1992).}
\nref\CZT{T.~Chen, R.N.~Zitter and R.~Tao, Southern Illinois University
Preprint, (1992).}
suggest that at fixed temperature there
are actually two critical electric fields, $E_{\tiny C1}$ and $E_{\tiny C2}$
($E_{\tiny C2}<E_{\tiny C1}$).
Below $E_{\tiny C2}$ the fluid consists primarily of a gas of monomers.
Above $E_{\tiny C1}$ the colloidal particles
clump together and phase separate into ``columns'' and dilute chains
as Halsey and Toor's argument would
suggest.  The columns may in fact be crystalline \refs{\T,\CZT}.
Between $E_{\tiny C2}$ and $E_{\tiny C1}$ there are
dense strings of colloidal spheres with random positions in any
constant-$z$ cross section.  The
simulation suggests a first order phase transition from
the random line phase to the clumped phase.  At lower densities,
experimental measurements do not suggest crystalline
order in the columns \ref\QUEST{
A.P.~Gast and C.F.~Zukoski, Advances in Colloid and Interface Sci. {\bf 30}:
153 (1989).}.
Thus, at low enough densities we can have coexistence of liquid-like
columns and a gas of chains, possibly terminating in a critical
point.

One shortcoming of our model in its present form
is that the chains it describes must start and
stop at the boundaries of the sample.
Both directed polymers and colloidal chains should be allowed to have
free ends within the sample.  As discussed in Section 4, free
ends may be incorporated by relaxing the constraint \eII\ and using
the free energy \refs{\LN-\N}
\eqn\eX{\eqalign{
F&=\int dz d^2\!x\cr&\left[{h\over 2}{\tb}^2 + {e\over 2}
\left(\grad\dr\right)^2 +{\alpha\over 2}\left(\dz\dr\right)^2
+{b\over 2}\dr^2 + {c\over 4!}\dr^4 +{G\over 2}(\dz\dr
+\grad\!\ddot\tb)^2\right]\cr}}
By taking $G\rightarrow\infty$, we recover our original model.  Additionally,
it can be shown that $G\sim \ell^{-1}$ as $\ell\rightarrow\infty$,
where $\ell$ is the chain length \refs{\LN,\KLN}.
The crossover from elongated objects to point objects when the chains are
finite will be discussed in detail in Section 4.

\subsec{Phase Separation in the Flux Liquid Phase}

Flux lines in superconductors may also be described by
an analogy with bosons \NS .  However, in the absence of
magnetic monopoles, flux lines cannot have free ends
inside the sample.  Their behavior, then, is much closer to
the behavior quantum mechanical bosons, as the constraint \eII\
is always satisfied.

Calculations to date for fluctuating flux lines have assumed a purely
repulsive pair potential \NS .  Muzikar and Pethicke
\ref\MP{P.~Muzikar and C.~Pethicke, Phys. Rev. B {\bf 24}: 2533 (1981).},
however, have argued that an effective attractive interaction arises
under some circumstances near $H_{\tiny C1}$ in the extreme type II limit.
This interaction would convert the continuous onset of flux penetration
at $H_{\tiny C1}$ to a first order transition.  An even more intriguing
situation would arise if the interaction between the flux lines had
{\sl two} distinct minima (see the inset to Figure 3).  In this case,
there should be a first order transition at $H_{\tiny C1}$, and a line
separating {\sl two} coexisting flux liquid phases, as in
Figure 2.  Our theory applies to the critical point marking the terminus
of this first order phase boundary.  The critical isochore is shown as
a dotted line in the figure.
\thefig{2}{Speculative phase diagram for Type II superconductors
with an attractive interaction between flux lines which
allows for coexisting dense and dilute entangled flux liquids.  At
sufficiently high temperatures, this line terminates in a
critical point.  The critical isochore, which marks the extension
of this line into the single phase region, is shown as
a dotted line.}
\thefig{3}{Pressure-temperature phase diagram for a quantum fluid
with the unusual pair potential shown in the inset.  The outer minimum
dominates in the superfluid phase $S1$, while the inner
minimum is primarily occupied in the higher pressure superfluid $S2$.  Phases
$S1$ and $S2$ are separated from a normal liquid $N$ by the dashed
$\lambda$-line.
Points $a$ and $b$ are conventional critical points.  By enhancing quantum
fluctuations at $T=0$ (say, by increasing $\hbar$), on arrives at the
critical point $c$ discussed in the paper.}
\subsec{Quantum Critical Phenomena}

Our analysis of course, can also be applied directly to
boson systems at $T=0$, which are equivalent to infinitely long
directed polymers \FEY .  In the case of a single particle species, we
would be describing the terminus of a line of first order transitions
separating superfluids coexisting at two different densities.  How this
situation might come about is illustrated in Figure 3.  The inset shows
an unusual pair potential with {\sl two} distinct minima describing
particles which, like helium, are light enough to avoid
crystallization at $T=0$ at low pressures.  The two minima lead to two
distinct liquid phases with increasing pressure.  At low pressures, the
stable zero temperature liquid has near neighbors occupying the deeper
outermost minimum.  As pressure is increased, there is a first
order phase transition to a denser liquid with the innermost minimum
occupied, followed eventually by crystallization at even higher pressure.
The two liquids must be superfluids at $T=0$.  In Figure 3, the parameters
have been adjusted so that the critical point for the dilute and dense
superfluids $S1$ and $S2$ occurs below the dashed $\lambda$-line describing
the transition to a normal liquid with increasing temperature.  There is a
second critical point associated with the coexistence of this normal
liquid with a dilute gas.

To obtain the critical point of interest to us here, we must increase
$\hbar$ (or decrease the mass) at $T=0$ until the two superfluid liquid
phases become indistinguishable as at the point $c$ in the figure.  Although
this {\sl gedanken} experiment is clearly rather esoteric, some control
over quantum fluctuations may be achievable for two-dimensional helium
films, where out of plane fluctuations alter the effective two-body
potential.  By varying the type of substrate, it may be possible to create
a purely repulsive effective potential and a novel superfluid gas phase
at $T=0$ \ref\COLE{E.~Cheng, M.W.~Cole and P.B.~Shaw,
J. Low Temp. Phys. {\bf 79}: 49 (1991); see, however, recent
work by C.~Carrero and M.~Cole for evidence {\sl against} this hypothesis.}.
We should stress, however, that we do not expect the unusual phase
diagram of Figure 3 for {\sl either} purely repulsive potentials or
conventional helium potentials with a single minimum.

The density operator of the bosons described above is given by
\edensx.
Upon differentiating with respect to $z$ we have
\eqn\edensi{
\dz\rho(\rb,z)=-\sum_i{d\rb_i(z)\over dz}\cdot\nabla_{{\tiny\perp},\rb}
\delta^2[\rb-\rb_i(z)]=-\grad\!\cdot\sum_i{d\rb_i(z)\over dz}\delta^2[\rb-
\rb_i(z)]}
The final sum on the right hand side is just \emomx , so we
have derived \eII. The quantity
$\tb(\rb,z)=\sum_i{d\rb_i(z)\over dz}\delta^2[\rb-
\rb_i(z)]$ is just the momentum operator of the bosons \refs{\NS,\KLN}.
On a more formal level, note that the ``time dependence'' ({\sl i.e.} the
$z$ dependence) of the quantum operator \edensx\ is given by the (Wick
rotated) Heisenberg equation of motion,
\eqn\eham{\dz\rho(\rb,z)=[\rho(\rb,z),{\cal H}],}
where ${\cal H}$ is the underlying quantum Hamiltonian.  Upon identifying
\eI\ with the integral over imaginary time of the
coarse-grained hydrodynamic Lagrangian associated with
this Hamiltonian (as is appropriate
for finite temperature quantum statistical mechanics), we
recover \eII\ from \eham\ upon using the commutation relation
between $\rho(\rb,z)$ and $\tb(\rb,z)$ \nref\F{
D.~Forster, {\sl Hydrodynamic Fluctuations, Broken Symmetry and
Correlation Functions}, (Benjamin/Cummings, Reading, MA, 1975); see also
the treatment of bosons in pp. 6-8
of A.A.~Abrikosov, L.P.~Gorkov and I.E.~Dzyaloshinski, {\sl Methods
of Quantum Field Theory in Statistical Physics}
(Dover, New York, 1963).},
\eqn\ecom{
[\rho(\rb,z),\tb(\rb',z)]={\kbt\over h}\grad\delta^2(\rb-\rb').}
In the hydrodynamic limit considered here, this commutator becomes
the classical ``Poisson bracket'' relation between density and momentum,
which arises because the momentum operator is the
generator of translations \F.  The constraint \eII\ thus directly reflects the
non-commutivity of the underlying position and momentum operators.

Note that our results represent a new universality class for Ising-like
quantum critical phenomena.  A related type of quantum
critical phenomena, extensively
studied in the past \ref\Y{
A.P. Young, J. Phys. C {\bf 8}: L309 (1975) and references therein.},
arises for the Ising model in a transverse magnetic field. The quantum
Hamiltonian, defined on a hypercubical lattice in $d$-dimensions is
\eqn\enwe{{\cal H}=-J\sum_{\langle i,j\rangle}\sigma_i^z\sigma_j^z -
h\sum_j\sigma_j^x}
where the $\{\vec\sigma_i\}$ are Pauli spin matrices.  The ordered state at
$T=0$ is destroyed for sufficiently large $h$.  This disordering transition
is known to be in the universality class of the $(d+1)$-dimensional classical
Ising model \Y .  Equation \enwe\ describes a quantum lattice gas, with the
world lines of the ``particles'' (up spins, say) in a $T=0$ path integral
formulation representing the directed polymers discussed in this paper.
Because the $\sigma_j^x$ is a spin flip operator, these ``polymers'' stop
and start at random in $(d+1)$-dimensions.  The appropriate coarse-grained
Landau description would be similar to \eI\ {\sl without}, however,
the constraint \eII .  The spin flip operator in \enwe\ does not respect
this conservation law, so it is not surprising that we find a different
universality class for boson critical phenomena.  Introducing free ends in
our model has the same effect as the spin flip operator in \enwe .  As
discussed above, we then find a crossover from the $(d+1)$-dimensional
uniaxial dipolar Ising critical phenomena to the $(d+1)$-dimensional Ising
behavior expected for \enwe .

\subsec{Directed Polymer Blends and Quantum Binary Mixtures}
By extending the original model to allow for binary mixtures,
we can study the
consolute point of two different directed polymer species.
In particular, we can
determine the critical exponents near the demixing point
of directed polymer blends and compare them to
those for point particles.  Analogous phenomena should occur
in binary mixtures of two different superfluids.  Among the possible
candidates would be films of \hes\ and \hef\ or possibly, a superfluid
monolayer of spin aligned hydrogen mixed with superfluid \hef.
As the mass ratio, $m_1/m_2$, of the two species
is increased towards unity with $\hbar>0$, quantum fluctuations
should eventually cause the two superfluids to mix.  The same effect may
be achievable by varying the properties of the substrate or formally,
by increasing $\hbar$.
A phase diagram as a function of temperature, composition and $\hbar$
is shown in Figure 4.  Note the incomplete phase separation appearing
for nonzero $\hbar$.  The theory applies to the critical point
at $T=0$.  In Section 5 we will analyze such
binary mixtures in greater detail.  Although we shall concentrate on $2+1$
quantum systems, our results are easily adapted to $3+1$ dimensional
quantum problems which are well described by mean field theory (see Section
4).

\subsec{Outline}

In Section 2 we formulate the Landau theory that will be the focus
of this paper.  We follow the derivation developed in \NS\ for
flux lines, which leads to a coherent state functional integral.  From there
we describe the hydrodynamic theory with and without free
ends as in \KLN .  In appendix
A we present a similar derivation for two
\thefig{4}{Phase diagram for a binary mixture of two
superfluid species as a function of the concentration of species B,
temperature and
Planck's constant.  This paper addresses the nature of the critical
point which occurs at $T=0$.}
\noindent interacting polymer species.

In Section 3 we describe the correlation functions of this
theory which could be measured by X-ray or neutron diffraction experiments.
We also discuss the meaning of the direction dependent compressibility
which characterizes these directed line systems.

In Section 4 we study the critical behavior of our model via an expansion
in $2-\epsilon+1$ dimensions.  We reproduce the results of Larkin and
Khmel'nitski\u\i\ for $\epsilon=0$
when there are no free ends \LK , and find that free
ends are a relevant perturbation leading to the critical behavior of a
$(d+1)$-dimensional Ising model with short range interactions.

Finally, in Section 5 we discuss the critical mixing of two
polymer or quantum boson species.  Using the free energy derived
in appendix A, we show that, just as in the point-particle case,
the critical mixing of line-like objects is in the same universality
class as single component critical phenomena.

\newsec{Coherent State Functional Integrals and Landau Theory}
The derivation of \eI\ from a more microscopic free energy for
directed lines follows
the presentations in \NS\ and \KLN .  We only outline the main points here,
first reviewing the theory for chains which span the system.  Internal
free ends are then introduced by adding a source to the boson coherent state
field theory \refs{\LN,\KLN}.

We start from the following path integral
partition function for the $N$ chains,
\eqn\eIIIi{\eqalign{Z_N&={1\over N!}\int\prod_{i=1}^N [d\rb_i]\cr
&\exp\lt\{
-{1\over \kbt}\int_0^L dz\lt[\sum_{i=1}^N{g\over2}
\lt({d\rb_i\over dz}\rt)^2+\sum_{i<j}^N
V[\rb_i(z)-\rb_j(z)]\rt]\rt\}\cr}\;}
where we have restricted our attention to
interactions between lines through an equal ``time''
potential $V[\rb_i(z)-\rb_j(z)]$.

The boson field theory is derived in three steps.  One first
writes the transfer matrix associated with \eIIIi\ in terms of
a many-particle real-time Schr\"odinger equation for the
polymer positions, and then second quantizes this equation
using bosons.  Finally, we pass to a coherent state functional integral
representation of the second quantized formalism.

The grand canonical partition sum which results is
\eqn\eIIIviii{Z_{\rm gr}\equiv\sum_{N=0}^\infty e^{L\mu N/\kbt}Z_N=
\int [d\psi][d\psi^*]\,e^{-S[\psi^*,\psi]}\;,}
where $\psi(\rb,z)$ is a complex boson field.  The boson action $S$
reads
\eqn\eIIIix{
S=\int_0^L dz\int d^2\!r
\left[\eqalign{
&\psi^*(\rb,z)\lt(\dz-
D\grad^2-\bar\mu\rt)\psi(\rb,z)+\cr
&{1\over2}\int d^2r'\,\bar v(\rb-\rb')\vert
\psi(\rb,z)\vert ^2\vert \psi(\rb',z)\vert ^2
\cr}\right]}
where $D=\kbt/2g$, $\bar\mu =\mu/\kbt$ and $\bar v =V/\kbt$.
Standard manipulations show that the density of flux lines is
\eqn\eIIInext{
\rho(\rb,z)=\vert\psi(\rb,z)\vert^2\;.}

\subsec{Entangled Limit with Free Ends}
We now assume that all phases are dense enough to be entangled and expand
around
the mean density $\rho_0$ \nref\HWA{See also Terry Hwa, unpublished}.
For polymers of length $\ell$, entanglement means
$\sqrt{D\ell}>>\rho_0^{-1/2}$, while
for quantum systems, we require very low temperatures so that the
thermal de Broglie wavelength is large compared to the particle spacing.
We parameterize the field $\psi$ by
$\psi=\sqrt{\rho}e^{i\theta}$, a change of variables with
unit functional determinant, and find
from \eIIIix\ that,
\eqn\eBi{
S=\int dz d^2\!r\,\left[{D(\grad\rho)^2\over 4\rho} + D\rho(\grad\theta)^2
+i\rho\partial_z\theta -\bar\mu\rho +{v\over 2}\rho^2\right]}
Upon introducing an auxiliary field $\tb(\rb,z)$ conjugate to $\grad\theta
(\rb,z)$ via a Hubbard-Stratonovich
transformation, and integrating out $\theta(\rb,z)$,
we arrive at the partition sum \eIa\ \refs{\KLN,\HWA}.
See Appendix A for the analogous transformation applied to binary mixtures
of quantum bosons.

We can add a (Poisson distribution) of free ends
to the theory by adding to the action
\eIIIix\ a term of the form \LN
\eqn\eBxii{\eqalign{
\delta S&=-\eta\int dzd^2\!r\,\left[\psi(\rb,z)+\psi^*(\rb,z)\right]\cr
&=-\int dz d^2\!r\, 2\eta\sqrt{\rho}\cos\theta\cr}}
The mean polymer length $\ell$ is related to $\eta$ by
$2\eta\ell\kbt=\sqrt{\rho_0}$ \KLN.
We may expand this term in powers of $\theta$ if the fluctuations
in $\theta$ are small, which they will be if the polymers are
sufficiently dense and entangled.
To leading order in $\theta$, equation \eBxii\ becomes
\eqn\eBxiii{\delta S\approx\hbox{const.}
+\int dzd^2\!r\,\eta\sqrt{\rho}\theta^2}
The Hubbard-Stratonovich transformation now leads to
\eqn\eBxv{
S=\int dzd^2\!r \left\{
{D\tb^2\over\rho} + {D(\grad\rho)^2\over 4\rho}
-\bar\mu\rho +{v\over 2}\rho^2 + {1\over
4\eta\sqrt{\rho}}\left[\partial_z\rho+\grad\!\ddot
\tb\right]^2
\right\}}
Upon expanding around an average density $\rho_0$, we arrive at
the theory for finite length polymers \eX , with $G=1/4\eta\sqrt{\rho_0}$ and
$h=2D/\rho_0$.

\newsec{Correlations and Compressibility}

{}From the free energy \eX\ we can, by considering only quadratic
terms, easily find the hydrodynamic form of the
structure function which could be
measured by either neutron or x-ray diffraction.
Here and henceforth we change units so that $e=1$.
The Fourier-transformed density-density correlation function is then
\eqn\eA{\eqalign{
S(\qp,\qz)&={\langle\dr(\qp,\qz)\dr(-\qp,-\qz)\rangle\over\rho_0^2}\cr
&={\kbt\left(h+G\qp^2\right)\over\left(G\alpha \qp^2+Gh+\alpha h\right)\qz^2 +
\left(Gb+h\right)\qp^2 + G\qp^4+hb}.
\cr}}
Note that $G$ drops out of \eA\ upon setting $h=0$.
This would have to be the case in the boson analogy, since
turning $h=0$ neglects quantum mechanics by decoupling the momentum
from the position.  The constraint \eII\
is unimportant in this classical limit.

In the limit $G\rightarrow\infty$ , we recover the correlations
typical of infinite polymers with no heads or tails from \eA ,
\eqn\eB{S(\qp,\qz)={\kbt\qp^2\over \left(h+\alpha\qp^2\right)\qz^2+b\qp^2
+\qp^4}.}
For small momenta, we can examine the curves of constant $S(\qp,\qz)$.  It is
easy
to see that these curves will go into the origin on straight lines, {\sl i.e.}
$\qz\sim\qp$.  This is a generic feature expected to hold for all
systems with the morphology of directed polymers, where the direction
is explicitly chosen by an external field.
Recalling that the limit as $\vert q\vert
\rightarrow 0$ of the structure function is related to the compressibility
of the sample \ref\PATH{
See for instance, R.K.~Pathria, {\sl Statistical Mechanics}, (Pergamon,
Exeter, Great Britain, 1972).}, we can determine the compressibility of a
sample of
directed lines, which will depend on the direction of the compression
\ref\rWIT{We thank Tom Witten for conversations on this point.}.
Upon considering a compressional wave at an angle $\phi$ from the $z$ axis,
we take the momentum to be $\qz=q\cos\phi$ and $\qp=q\sin\phi$,
and find for small $q$,
\eqn\eC{S(\qp,\qz)\mathrel{\mathop\sim_{q\rightarrow 0}}
{\kbt\over b+h\left(\cot\phi\right)^2}=\kbt\chi_{\tiny T}(\phi),}
where $\chi_{\tiny T}$ is the compressibility, and is a function of the
direction $\phi$ in which the compression is made.
Note again that the $h=0$ limit
is especially simple:  The compressibility
no longer depends on the compression angle if tangent fluctuations
are ignored.  We then recover the result for an isotropic
fluid of point particles, namely $\chi_{\tiny T}=b^{-1}$.
The compressibility enters the linear response relation:
\eqn\elin{\langle\dr(\qp,\qz)\rangle
=\chi(\qp,\qz)\delta\sigma(\qp,\qz),}
where $\delta\sigma$ is a sinusoidally varying force directed along
${\vec q}=(\qp,\qz)$.
The unusual $\phi$-dependence of this
response function may be understood as follows:
Equations \elin\ and \eC\ predict that $\langle\dr\rangle=0$ for compressions
when $\phi\rightarrow 0$.  This result arises because squeezing the system
along the $z$-axis will not change
the number of polymers piercing any constant $z$-slice.  In general, the
linear response \elin\ depends on the directions of the compression relative
to $\hat z$,
as parameterized by $\phi$.

Returning to finite length polymers, and hence to finite $G$, we repeat
the above analysis and find a $\phi$-independent result
\eqn\eD{S(\qp,\qz)\mathrel{\mathop\sim_{q\rightarrow 0}}
{\kbt\over b}.}
Because at sufficiently large distances (larger than the average
polymer length $\ell$) finite length polymers will
ultimately behave like point particles,
the long-wavelength properties of the sample will
not exhibit any character of the directed constituents.  At intermediate
scales, the factors of $\bo{G\qp^2}$ may not be neglected in \eA , and
we return to the direction dependent result \eC .

\newsec{Renormalization Group and Finite Length Polymers}

In the last section we saw that at long-wavelengths the line-like
nature of the directed melt disappears for finite chain lengths.
Here we construct a renormalization group which allows us to
study this crossover in detail.  When $G\rightarrow\infty$, we
recover the results of Larkin and Khmel'nitski\u\i\ \LK .  For $G$ finite,
however, the system crosses over to the behavior of a $(d+1)$-dimensional
Ising model with short range interactions \Y .
We shall treat the $z$-direction
separately from the directions in the plane perpendicular to $z$.
We let $d$ be the dimension of the {\sl bosons} so that the
corresponding directed polymers exist in $(d+1)$-dimensions (see Figure 1).

Consider first the graph which renormalizes the four-point
interaction $\dr^4$.  To interpret this graph we
need to invert a $(d+1)\times (d+1)$ matrix to get the
propagator, as there are $d$ degrees of freedom in $\tb$ and an additional
degree of freedom from the density fluctuations, $\dr$.  However, as mentioned
earlier, the transverse modes of $\tb$ decouple.  To make this
explicit, we decompose $\tb$ in momentum space,
\eqn\etdec{
\tb(\qp,\qz)=\tb_L+\tb_T,} with
\eqn\etdeca{t_{L,j}={q_jq_k\over \qp^2}t_k} and
\eqn\etdecb{t_{T,j}=\left(\delta_{jk}-{q_jq_k\over\qp^2}\right)t_k}
In \etdeca\ and \etdecb\ the indices
$j$ and $k$ only run over the $d$-dimensional space perpendicular
to $z$ and we use the summation convention.  Additionally we take
$t_{L,j}=-iq_j\pi(\qp,\qz)$.  With this parameterization,
the fourier transform of the divergence of $\tb$ is simply
$i{\bf q}_{\tiny\perp}\cdot\tb=\qp^2\pi$, and the fourier transform of
the $\tb^2$ term in \eX\ is $\int d\qz d^d\!\qp\,[\qp^2\pi^2+{\tb_T}^2-
{1\over\qp^2}(\qp\cdot{\tb_T})^2]$.  Thus we can integrate out $\tb_T$
from \eX , and we are left with a theory only involving $\dr$ and $\pi$.

Inverting the $2\times 2$ quadratic form which remains gives the following
useful propagators:
\eqna\eprop{$$\eqalignno{
\langle\vert\dr(\qp,\qz)\vert^2\rangle&={
h+G\qp^2\over\left(G\alpha \qp^2+Gh+\alpha h\right)\qz^2 +
\left(Gb+h\right)\qp^2 + G\qp^4+hb}\qquad&\eprop a\cr
\langle\vert\pi(\qp,\qz)\vert^2\rangle&={1\over\qp^2}{
\qp^2+b+\alpha\qz^2+G\qz^2\over\left(G\alpha \qp^2+Gh+\alpha h\right)\qz^2 +
\left(Gb+h\right)\qp^2 + G\qp^4+hb}\qquad&\eprop b\cr
\langle\dr^*(\qp,\qz)\pi(\qp,\qz)\rangle&={iG\qz\over
\left(G\alpha \qp^2+Gh+\alpha h\right)\qz^2 +
\left(Gb+h\right)\qp^2 + G\qp^4+hb}\qquad&\eprop c\cr
}$$}
After inserting these propagators into the graph shown in Figure 5,
we integrate out the
frequency-like variable $\qz$ and find the renormalized four point interaction
associated with equation \eX\ is $c_{\tiny R}=c+\delta c+\bo{c^3}$, where
\eqn\edeltac{
\delta c=-{3c^2\over 8}{2\over\Gamma ({d\over 2})(4\pi )^{d/2}}
\int {\qp^{d-1}d\qp\over \qp^2}
{\sqrt{1+{\displaystyle{h\over G\qp^2}}}\over
\sqrt{h+\alpha\qp^2+\alpha{h\over G}}\sqrt{\left(1+{\displaystyle{b\over
\qp^2}}\right)^3}}}
Near the critical point, $b\approx 0$ and so when $G=\infty$,
$\delta c$ diverges in the infrared for $d\le 2$.  This
infrared divergence suggests an $\epsilon$-expansion around $d=2$, where
$\epsilon=2-d$.
Our approach follows
\ref\rFSN{D.~Forster, D.R.~Nelson, and M.J.~Stephen, Phys. Rev. A {\bf 16}: 732
(1977).},
which is a variation of the dynamic renormalization
group of
\ref\rHH{B.I.~Halperin, P.C.~Hohenberg, and S.K.~Ma,
Phys. Rev. Lett. {\bf 29}: 1548 (1972).}.
First we integrate out a momentum shell in $\qp$
from $\Lambda e^{-\ell}$ to $\Lambda$, but integrate freely over $\qz$.
We then rescale our variables so that the ultraviolet
cutoff in the perpendicular direction
is held fixed. After rescaling $\rho$ and $\tb$ accordingly,
we are left with the same theory but with different coupling constants.
When this procedure is iterated, $c$ is driven toward a fixed
point which describes the universal long wavelength critical behavior
for large $G$.

\subsec{Momentum Shell Integration}
We must first integrate out the transverse momentum in the
range $\Lambda e^{-\ell}<\qp<\Lambda$. This can be done straightforwardly
by expanding the functional integral in $c$.  Care must be taken to
account for all possible contractions of the operators in the
expansion.  The symmetry factors can be found in the usual way for
Wick expansions.  It is important to note that diagrams
renormalize the remaining low
\thefig{5}{Hartree graph which diverges for $d<2$.  We integrate here
over all internal momenta.}
\noindent momentum modes, and are not
simply expectation values. Graphs which renormalize the
quadratic and quartic contributions to Landau theory are shown
in Figure 6.
Upon carrying out this procedure, we arrive at the following relations for the
intermediate values of the coupling constants in \eX
\eqna\eINT{$$\eqalignno{
h'&=h&\eINT a\cr
\alpha'&=\alpha&\eINT b\cr
b'&=b+{c\over 4\sqrt{G}}\int_{\Lambda e^{-\ell}}^\Lambda {d^d\!\qp\over (2\pi
)^d}\,
\left[{\sqrt{h+G\qp^2}\over\sqrt{b+\qp^2}}{1\over\sqrt{h+\alpha\qp^2+\alpha{
{h\over G}}}}
\right]&\eINT c\cr
c'&=c\left(1-{3c\over 8\sqrt{G}}
\int_{\Lambda e^{-\ell}}^\Lambda {d^d\!\qp\over (2\pi)^d}\,
\left[{\sqrt{h+G\qp^2}\over\sqrt{(b+\qp^2)^3}}
{1\over\sqrt{h+\alpha\qp^2+\alpha{
{h\over G}}}}
\right]\right)&\eINT d\cr
G'&=G&\eINT e\cr}$$}
where we have gone to one-loop order.
In \eINT{}\ we have already
integrated over $\qz$.

We rescale
perpendicular lengths by $R_{\tiny\perp}'=R_{\tiny\perp}e^{-\ell}$ and the $z$
direction
by $Z'=Ze^{-\int_0^\ell \gamma(\ell')d\ell'}$.
In momentum space, these rescalings read
$\qp'=\qp e^{\ell}$ and $q_z'=q_z e^{\int_0^\ell \gamma(\ell')d\ell'}$,
where the function $\gamma\ofl$ is to be determined.

When doing the first momentum
shell integration, the coupling constants were
independent of length scale.  However, they then acquire a momentum
dependence because we have absorbed the large momentum
effects into them. The correct renormalized theory is a coupled set
of integral equations where the coupling constants are taken
to be scale dependent.  An alternative, but equivalent approach
is to integrate over a small momentum range where the coupling constants
are approximately fixed and then repeat the entire calculation iteratively.
This leads to the usual differential renormalization group
equations.

\subsec{Recursion Relations and Critical Behavior}
We now choose an infinitesimal momentum shell $e^{-\delta}$, and take the
limit $\delta\rightarrow 0$. This leads to
differential renormalization
group equations which can be integrated
\thefig{6}{Graphs which contribute
to the renormalization group calculation.  In these graphs, the internal
lines are integrated over all values of $\qz$ but $\qp$ is only integrated
in a momentum shell between $\Lambda e^{-\ell}$ and $\Lambda$.
Figure (a) is a graph which renormalizes the propagator, and figure
(b) is a graph which renormalizes the four-point coupling $c$.}
\noindent to produce the couplings appropriate
for a cutoff $\Lambda e^{-\ell}$ in the perpendicular
direction.  We use units such that $\Lambda=1$ in
the following.  We also define the renormalized ``reduced temperature''
\eqn\ered{r=b+{c\over 2\sqrt{h+\alpha}}A_d,}
which vanishes at the critical point in the fluctuation-corrected theory.
The constant $A_d = 2/[\Gamma ({d\over 2})(4\pi )^{d/2}]$
is a geometrical factor.
The differential recursion relations valid near $d=2$ are

\eqna\eINTi{$$\eqalignno{
\der{h}&=h\left(4-2\gamma \right)&\eINTi a\cr
\der{\alpha}&=\alpha\left(2-2\gamma\right)&\eINTi b\cr
\der{r}&=r\left(2-{c\over 16\pi\sqrt{h+\alpha +\alpha{
{h\over G}}}}\right)
+{ch\over 16\pi G\sqrt{h+\alpha+\alpha{{h\over G}}}}
&\eINTi c\cr
\der{c}&=c\left(4-d-\gamma-{3c\over 16\pi\sqrt{h+\alpha+\alpha{
{h\over G}}}}
\right)&\eINTi d\cr
\der{G}&=G\left(2-2\gamma\right)&\eINTi e\cr}$$}

We set $\gamma=2$ in order to hold $h$ fixed, and note that $\alpha$ is
an irrelevant variable.  We have expanded about $G^{-1}=0$.
Examining these recursion relations, we see that perturbation theory
amounts to an expansion in two parameters, namely
$\bar c=c/\sqrt{h+\alpha{h\over G}}$
and $(\bar G)^{-1}=hG^{-1}$.  These couplings have the recursion relations
\eqna\eINTii{$$\eqalignno{
&{d\bar c\over d\ell}=\bar c\left(\epsilon-{3\bar c\over 16\pi}\right)&\eINTii
a\cr
&{d\over d\ell}\left({h\over G}\right)=2\left({h\over G}\right)&\eINTii
b\cr}$$}
where $\epsilon=2-d$.

We first consider the subspace of theories with $G=\infty$.  In this case
there is a stable, nontrivial fixed point for positive $\epsilon$.  Though
we have already noted that this theory is identical to the uniaxial
ferroelectric, and hence should have all the same critical behavior,
a direct calculation of these exponents provides a useful
check of our renormalization group method.
In particular we can calculate the logarithmic
corrections to the compressibility discussed in Section 3 in the
critical dimension of $2+1=3$.
Upon setting $\epsilon=0$ in \eINTii{a}\ we solve for $\bar c\ofl$
and find that
\eqn\ebarc{
\bar c\ofl={\bar c(0)\over 1+{\displaystyle{3\bar c(0)\over 16\pi}}\ell},}
which leads via \eINTi{c}\ to
\eqn\ebofl{
r\ofl=r(0)e^{2\ell}\left(1+{\displaystyle{3\bar c(0)\over
16\pi}}\ell\right)^{-1/3}}
Since at this order in $\epsilon$ there is no nontrivial rescaling
of the fields, we have
\eqn\esusc{\eqalign{
\chi(\qp&,\qz;r_0)=\int dzd^d\!x\,e^{i\qp\cdot {\bf x}-i\qz z}\langle\rho({\bf
x},z)\rho({\bf 0},0)\rangle_{\ell=0}\cr
&=\int e^{\int_0^\ell\gamma(\ell')d\ell'}dz'e^{d\ell}d^d\!x'\,\cr
&\qquad\times\left(e^{-\half\int_0^\ell\gamma(\ell')d\ell'}
e^{(2-d)\ell/2}\right)^2e^{i\qp'\cdot {\bf x}'-i\qz' z'}\langle\rho'({\bf
x}',z')\rho'({\bf 0},0)\rangle_\ell\cr
&=e^{2\ell}\chi_\ell\left(e^{\ell}\qp,
e^{\int_0^\ell\gamma(\ell')d\ell'}\qz;r\ofl\right)
\cr}}
We now choose $\ell=\ell^*$ large enough so that $r(\ell^*)$ has grown to
$\bo{1}$.
Then we have from \ebofl\
\eqn\eboflagain{1=r(\ell^*)
=r_0(3\bar c_0/16\pi)^{-1/3}e^{2\ell^*}(\ell^*)^{-1/3}}
so that
\eqn\enext{e^{2\ell^*}\sim r_0^{-1}\vert\ln r_0\vert^{1/3}}
and thus
\eqn\esusce{\chi(\qp,\qz;r_0)={\vert\ln r_0\vert^{1/3}\over
r_0}\chi(\xi_{\tiny\perp}\qp,
\xi_z\qz;1)}
where $\xi^2_{\tiny\perp}\sim\xi_z\sim(\ln r_0)^{1/3}/r_0$,
in agreement with the results of \nref\BZJ{E.~Br\'ezin and
J.~Zinn-Justin, Phys. Rev. B {\bf 13}: 251 (1976).}\LK .
Note the anisotropically diverging correlation lengths produced
by the directed line-like nature of the degrees of freedom.  The two-loop
calculation of \BZJ\ leads to the result $\xi_z\sim\xi_{\tiny\perp}
^{2-{2\epsilon^2\over 243}}$ to $\bo{\epsilon^2}$.
Recalling the discussion in the last section of the compressibility, we have
finally
\eqn\edirren{\chi(q\sin\phi,q\cos\phi;r_0)
\mathrel{\mathop\sim_{q\rightarrow 0}}{\vert\ln r_0\vert^{1/3}\over r_0}
{1\over 1+h(\vert\ln r_0\vert^{1/3}/r_0)
\left(\cot\phi\right)^2}}
which should be compared to the mean field result \eC .
Note that the effective direction of the compressional wave changes as we
go to longer length scales.  Near the critical point we always find
$\chi(\phi)\sim
{1\over h\left(\cot\phi\right)^2}$, which appears to be a universal result.

Similarly, in the ordered phase, we can calculate the
specific heat exponent by flowing along a renormalization group
trajectory and matching on to an effective free energy at long-wavelengths
(A more complicated calculation shows that this behavior occurs above
$T_{\tiny C}$ as well).
In the ordered phase, the bulk free energy is approximately $-3\Omega(r^2/2c)$,
where $\Omega$ is the total volume of the system.  We then have
\eqn\efree{\eqalign{
F_\ell&=-3\Omega_\ell{r(\ell)^2\over 2c\ofl}\cr
&=-{3\over
2}e^{-d\ell}e^{-\int_0^\ell\gamma(\ell')d\ell'}\Omega_0{r(\ell)^2\over
c\ofl}\cr}}
Again choosing $\ell=\ell^*$ as in \eboflagain\ we find
that
\eqn\efreeatlast{F_\ell=-{3\over2}e^{-d\ell^*}e^{-\int_0^{\ell^*}
\gamma(\ell')d\ell'}\Omega_0{1^2\over c(\ell^*)}\sim e^{-(d+2)\ell^*}\ell^*
\sim r_0^2\vert\ln r_0\vert^{1/3}}
Differentiating twice with respect to $r_0$, we recover the correct
specific heat-like behavior at the critical point.

Going back to the full theory we now allow $G^{-1}$ to take on finite values.
In this case we are driven away from our fixed point and $G^{-1}$ flows towards
$\infty$.  This reflects a crossover to an anisotropic but {\sl point-like}
phase where
the constraint \eII , and hence the line-like nature
of the polymers is unimportant.  Because $\tb$ decouples from the
density fluctuations the quantum mechanical bosons behave classically at
long wavelengths.
A similar crossover occurs in polymer nematics when hairpins are
introduced into the theory and the previously directed polymers become
isotropic \KLN .

\newsec{Mixing Exponents}

We now consider directed polymer blends, or, equivalently, binary
mixtures of two superfluids.
As with binary mixing in classical systems of point particles, the mixing
exponents are identical to the liquid-gas exponents.  We demonstrate
this equivalence by showing that the Landau-Ginzburg theory controlling
the mixing fraction is the same as that for a single component critical
point, \eI .  For simplicity, we restrict our attention to the limit
$G\rightarrow\infty$.
In appendix A we derive the quadratic part of the free
energy of two polymer species using the boson representation.
To these terms we add nonlinear couplings to
account for the configurational entropy of each polymer species, multiparticle
interactions, {\sl etc.}, and
consider the model free energy
\eqn\emI{F=\int dz d^d\!x\,\left[\eqalign{
&{h_1\over 2}\tb_1^2+{h_2\over 2}\tb_2^2 + h_{12}\tb_1\!\cdot\!\tb_2
+\half(\grad\dr_1)^2+\half(\grad\dr_2)^2\cr
&+{b_1\over 2}\dr_1^2+{b_2\over 2}\dr_2^2
+b_{12}\dr_1\dr_2 + {w_1\over 3!}\dr_1^3+{w_2\over 3!}\dr_2^3 \cr
&+ {c_1\over 4!}
\dr_1^4+{c_2\over 4!}\dr_2^4 +{c_{12}\over 4}\dr_1^2\dr_2^2\cr}\right]}
subject to the constraints
\eqn\emII{\dz\dr_j+\grad\ddot\tb_j=0\qquad{j=1,2}.}
We have neglected gradient terms of the form $(\dz\dr_i)^2$, because
these turn out to be irrelevant at the demixing point.
Terms proportional to $\dr_1\dr_2^2$ and $\dr_2\dr_1^2$ could
have been added, but do not affector our results in any essential way.
Additionally, we assume that
the $(\grad\dr_i)^2$ terms have been
diagonalized so that the term $\grad\dr_1\cdot
\grad\dr_2$ is not present.
It is useful to pass to sum and difference variables,
{\sl i.e.} $\dr_{\tiny \pm}=(\dr_1\pm\dr_2)/\sqrt{2}$.
We only keep terms in the free energy up to second order in
the noncritical mode $\dr_{\tiny +}$, but keep all terms
up to quartic order in $\dr_{\tiny -}$.  We also define fields
$\tb_{\tiny \pm}=(\tb_1\pm\tb_2)/\sqrt{2}$ which are the tangents
associated with the
sum and difference variables.  The free energy now takes the form
\eqn\emIII{
F=\int dzd^d\!x\,\left[\eqalign{
&{h_{\tiny +}\over 2}\tb_{\tiny +}^2+{h_{\tiny -}\over 2}\tb_{\tiny -}^2
+h_{\tiny +-}\tb_{\tiny +}\tb_{\tiny -} + \half(\grad\dr_{\tiny +})^2
+\half(\grad\dr_{\tiny -})^2\cr
&+{b_{\tiny +}\over 2}\dr_{\tiny +}^2+{b_{\tiny -}\over 2}\dr_{\tiny -}^2
+ b_{\tiny +-}\dr_{\tiny +}\dr_{\tiny -}
+{w_{\tiny -}\over 3!}\dr_{\tiny -}^3 + {c_{\tiny -}\over 4!}\dr_{\tiny
-}^4\cr}\right]}
with the constraints now reading
\eqn\emIV{\dz\dr_{\tiny \pm}+\grad\ddot\tb_{\tiny \pm}=0.}
We can again integrate out the transverse parts of $\tb_{\tiny\pm}$, and
use the constraint \emIV\ to solve for the longitudinal parts.
The resulting quadratic part of the free
energy reads (in momentum space)
\eqn\emV{F_{\hbox{quad}}=\int {d\qz\over 2\pi}{d^d\!\qp\over (2\pi)^d}\,
\left[\eqalign{
&\half\left(h_{\tiny +}{\qz^2\over\qp^2}+\qp^2+b_{\tiny
+}\right)\vert\dr_{\tiny +}\vert^2\cr
&\qquad+\half\left(h_{\tiny -}{\qz^2\over\qp^2}+\qp^2+b_{\tiny
-}\right)\vert\dr_{\tiny -}\vert^2\cr
&\qquad\qquad+h_{\tiny +-}{\qz^2\over\qp^2}\dr_{\tiny +}\dr_{\tiny -}
+ b_{\tiny +-}\dr_{\tiny +}\dr_{\tiny -}\cr}\right]}
At this point we are in the position to integrate out $\dr_{\tiny +}$,
resulting in
\eqn\emVI{\eqalign{
&F_{\hbox{quad}}'[\dr_{\tiny -}]
=\cr&
\half\int {d\qz\over 2\pi}{d^d\!\qp\over (2\pi)^d}\,
\left[h_{\tiny -}{\qz^2\over\qp^2}+\qp^2+b_{\tiny -}-{\left(h_{\tiny +-}
{\displaystyle{\qz^2\over\qp^2}}+b_{\tiny +-}
\right)^2\over h_{\tiny +}{\displaystyle{\qz^2\over\qp^2}}+\qp^2+b_{\tiny +}}
\right]\vert\dr_-\vert^2\cr}.}
Upon expanding the denominator of \emVI\ in $\qp$ and $\qz$,
keeping only relevant and
marginally relevant terms, and reintroducing $\tb_{\tiny -}$ we finally have
\eqn\emVII{F'=\int dzd^d\!x\,\left[
\eqalign{
&\half\left(h_{\tiny -}-{2b_{\tiny +-}h_{\tiny +-}\over b_{\tiny +}}
+{b_{\tiny +-}^2h_{\tiny +}\over b_{\tiny +}^2}\right)\tb_{\tiny -}^2
+\half\left(1+{b_{\tiny +-}^2\over b_{\tiny +}^2}\right)(\grad\dr_{\tiny
-})^2\cr
&+\half\left(b_{\tiny -} - {b_{\tiny +-}^2\over b_{\tiny +}}\right)\dr_{\tiny
-}^2
+{w_{\tiny -}\over 3!}\dr_{\tiny -}^3 + {c_{\tiny -}\over 4!}\dr_{\tiny -}^4\cr
}\right]}
subject to
\eqn\emVIII{\dz\dr_{\tiny -}+\grad\ddot\tb_{\tiny -}=0.}

As in the single component model, the critical mixing line is gotten by
setting $w_{\tiny -}=0$.  This could be done by altering the chemical
potentials of the two species, for instance.  Our model is
thus equivalent to the original model \eI , except for the
unimportant coupling $\alpha(\dz\dr_{\tiny -})^2$. We conclude that the
critical mixing of {\sl line-like} objects is also in the universality class
of the three-dimensional uniaxial ferroelectric.

\newsec{Acknowledgments}

It is a pleasure to acknowledge helpful conversation with L.~Balents, P.~Clark,
M.~Cole, B.I.~Halperin,
T.~Hwa, P.~Le Doussal, G.~McKinley, R.B.~Meyer, O.~Narayan and T.~Witten
during the course of this investigation.
One of us (RDK), would like to acknowledge the support of a National Science
Foundation Graduate Fellowship.  This work was supported by the National
Science Foundation, through Grant DMR91-15491 and through the
Harvard Materials Research Laboratory.

\appendix{A}{Derivation of Hydrodynamics for Binary Mixtures}

Our starting point is a pair of boson field theories, each
describing a distinct superfluid or polymer species \NS . We then couple them
through their densities, as well as their ``currents'' or
tangent fields.  More precisely the
partition function is
\eqn\eAO{Z=\int[d\psi^*_1][d\psi_1][d\psi^*_2][d\psi_2]e^{-S[\psi^*_1,\psi_1,
\psi^*_2,\psi_2]}}
where the action is
\eqn\eAI{{F\over\kbt}\equiv S=S_{1}[\psi^*_1,\psi_1]+S_{2}[\psi^*_2,\psi_2]+
S_{12}[\psi^*_1,\psi_1,\psi^*_2,\psi_2].}
The individual terms read
\eqn\eAII
{S_j=\int dz d^d\!x\,\left[\psi_j^*\left(\dz - D_j\grad^2 -\bar\mu_j
\right)\psi_j + {B_j\over 4}\vert\psi_j\vert^4\right]}
while the interaction is
\eqn\eAIII{\eqalign{&S_{12}=\cr &\int
dzd^d\!x\,\left[B_{12}\vert\psi_{1}\vert^2
\vert\psi_{2}\vert^2 + \eta_{12}
\left(\psi^*_{1}\grad\psi_{1}-\psi_{1}\grad\psi^*_{1}
\right)\cdot
\left(\psi^*_{2}\grad\psi_{2}-\psi_{2}\grad\psi^*_{2}
\right)\right]\cr}}
The change of variables
\eqn\eAIV{\psi_j=\sqrt{\rho_j}e^{i\theta_j}}
leads to
\eqn\eAV{S=\int dz d^d\!x\,\left\{\eqalign{&
i\rho_{1}\dz\theta_{1}
+{D_{1}(\grad\rho_{1})^2\over 4\rho_{1}}
+D_1\rho_{1}(\grad\theta_{1})^2
-\bar\mu_{1}\rho_{1}+{B_1\over 4}\rho_{1}^2\cr
&i\rho_{2}\dz\theta_{2}
+{D_{2}(\grad\rho_{2})^2\over 4\rho_{2}}
+D_2\rho_{2}(\grad\theta_{2})^2
-\bar\mu_{2}\rho_{2}+{B_{2}\over 4}
\rho_{2}^2\cr
&+B_{12}\rho_{1}\rho_{2} +\eta_{12}\rho_{1}
\rho_{2}\grad\theta_{1}
\cdot\grad\theta_{2}\cr}\right\}}
We now introduce two fields which we shall see are proportional to the tangent
fields of the individual species.  We start with the identity
\eqn\eAVI{\eqalign{
&\exp\left\{\int dz d^d\!x\,\left[ D_1\rho_{1}(\grad\theta_{1})^2
+D_2\rho_{2}(\grad\theta_{2})^2 + \eta_{12}\rho_{1}
\rho_{2}\grad\theta_{1}\cdot\grad\theta_{2}\right]\right\}
\cr&\quad=\int [d\pb_{1}][d\pb_{2}]
\quad\exp\left\{\int dzd^d\!x\,\left[\eqalign{
&{D_2\rho_2\pb_1^2 + D_1\rho_1\pb_2^2
-2\eta_{12}\rho_1\rho_2\pb_1\!\cdot\!\pb_2
\over D_1D_2\rho_1\rho_2 -\eta_{12}^2
\rho_1^2\rho_2^2}\cr & +
i\pb_1\!\cdot\!\grad\theta_1+i\pb_2\!\cdot\!\grad\theta_2\cr}\right]
\right\}}}
Upon using this identity to replace the terms quadratic in $\theta$ appearing
in \eAV\ and integrating over $\theta_j$, we find that
\eqn\eAVII{
Z=\int [d\rho_1][d\rho_2][d\pb_1][d\pb_2]\delta[\dz\rho_1+\grad\ddot\pb_1]
\delta[\dz\rho_2+\grad\ddot\pb_2]e^{-S'[\rho_1,\rho_2,\pb_1,\pb_2]}}
where
\eqn\eAVIII{
S'=\int dzd^d\!x\,\left\{\eqalign{&{D_2\rho_2\pb_1^2 + D_1\rho_1\pb_2^2
-2\eta_{12}\rho_1\rho_2\pb_1\!\cdot\!\pb_2
\over D_1D_2\rho_1\rho_2 -\eta_{12}^2
\rho_1^2\rho_2^2}\cr
&+ {D_1(\grad\rho_1)^2\over 4\rho_1} + {D_2(\grad\rho_2)^2\over 4\rho_2}\cr
& -\bar\mu_1\rho_1 -\bar\mu_2\rho_2 + {B_1\over 4}\rho_1^2+
{B_2\over 4}\rho_2^2
+B_{12}\rho_1\rho_2\cr}\right\}}
We now expand $\rho_i=\rho_{0,i} + \dr_i$ around the minimum of the
bulk free energy and rescale $\dr_j$ so that the kinetic term
in the action is ${1\over 2\kbt}(\grad\dr_j)^2$ to obtain
\eqn\eAIX{
S'={1\over\kbt}\int dz d^d\!x\,\left\{\eqalign{&
{h_1\over 2}\tb_1^2+{h_2\over 2}\tb_2^2+h_{12}\tb_1\!\cdot\!\tb_2
+{1\over 2}(\grad\dr_1)^2+\half(\grad\dr_2)^2\cr
&+{b_1\over 2}\dr_1^2+{b_2\over 2}\dr_2^2 + b_{12}\dr_1\dr_2\cr}\right\}}
where $\tb_j=\sqrt{D_j/2\rho_{j,0}}\pb_j$ and
the $h_j$ and $b_j$ can be read off from \eAVIII\ in the
expansion of $\rho_j$ around $\rho_{j,0}$.  The delta-functionals in \eAVII\
impose the constraints
\eqn\eAX{\dz\dr_j+\grad\ddot\tb_j=0.}

\listrefs

\listfigs
\bye